\documentclass[12pt,preprint]{aastex}
\usepackage{epsfig}

\def\gtrsim{\mathrel{\hbox{\rlap{\hbox{\lower4pt\hbox{$\sim$}}}\hbox{$>$}}}}
\def\lesssim{\mathrel{\hbox{\rlap{\hbox{\lower4pt\hbox{$\sim$}}}\hbox{$<$}}}}

\def\vkm{km s$^{-1}$}

\def\degree{$^\circ$}

\def\arcsa#1#2{$#1^{\prime\prime}_{^\textrm{.}}#2$}

\def\solarmass{$M_\odot$}

\def\msperyr{$M_\odot$ yr$^{-1}$}

\def\mJyb{mJy beam$^{-1}$}

\def\tlabel#1{(\textit{#1})}

\def\cmc{cm$^{-3}$}
\def\cmce{\textrm{cm}^{-3}}

\def\mH2{m_{\textrm{\scriptsize H}_2}}

\def\H2{H$_2$}
\def\N2HP{N$_2$H$^+$}

\def\NH3{NH$_3$}

\def\putfig#1#2#3{\epsfig{scale=#1,angle=#2,figure=#3}}
\def\putfiga#1#2#3{}
\def\leftblank#1{}

\newcounter{mfigure}[section]

\begin{document}

\title{Magneto-Centrifugal Origin for Protostellar Jets Validated through
Detection of Radial Flow at the Jet Base}

\author{Chin-Fei Lee$^{1,2}$, Zhi-Yun Li$^3$, Hsien Shang$^1$, \& Naomi Hirano$^1$}

\altaffiltext{1} {Academia Sinica Institute of Astronomy and Astrophysics, No.  1, Sec. 
  4, Roosevelt Road, Taipei 10617, Taiwan}
\altaffiltext{2} { Graduate Institute of Astronomy and Astrophysics, National Taiwan 
   University, No.  1, Sec.  4, Roosevelt Road, Taipei 10617, Taiwan}
\altaffiltext{3} {Astronomy Department, University of Virginia, Charlottesville, VA 22904, USA}


\begin{abstract} 
Jets can facilitate the mass accretion onto the protostars
in star formation.  They are believed to be launched from accretion
disks around the protostars by magneto-centrifugal force, as supported by
the detections of rotation and magnetic field in some of them. Here we
report a radial flow of the textbook-case protostellar jet HH 212 at
the base to further support this jet-launching scenario.  This radial
flow validates a central prediction of the magneto-centrifugal theory of
jet formation and collimation, namely, the jet is the densest part of a
wide-angle wind that flows radially outward at distances far from the
(small, sub-au) launching region.  Additional evidence for the radially 
flowing wide-angle component comes from its ability to reproduce the
structure and kinematics of the shells detected around the HH 212 jet.  This
component, which can transport material from inner to outer disk, could
account for the chondrules and Ca-Al-rich inclusions detected in the Solar
System at large distances.  
\end{abstract}


\section{Introduction}

Protostellar jets are seen associated with young stellar objects
\citep{Bally2016,Anglada2018,Lee2020Jet}.  They are ejected from the
innermost parts of accretion disks \citep{Ray2007} and linked to the process
of mass accretion onto the central protostars
\citep{Cabrit1990,Hartigan1990}.  However, how they are launched is still
under debate.  Recently, they are found to be spinning
\citep{Bacciotti2002,Lee2017Jet} and magnetized \citep{Lee2018BJet}, as
expected from magneto-centrifugal launching models, carrying angular
momentum away from the disks.  However, more observations of protostellar
jets are still needed to confirm and constrain the launching models.

HH 212 is a textbook-case protostellar jet because of its well-defined
structure with high collimation and symmetry \citep{Zinnecker1998}.  The jet
is ejected from a rotating accretion disk around a low-mass protostar
\citep{Codella2014,Lee2017COM} at a distance of $\sim$ 400 pc in Orion.  It
lies almost in the plane of the sky \citep{Claussen1998}.  It was recently
found to be rotating, carrying angular momentum away from the innermost part
of the disk \citep{Lee2017Jet}, consistent with the magneto-centrifugal
launching models.  Here, we search for another tell-tale sign to support the models
 --  a radial flow at the base of the jet.  This radial flow is
required because, in the classic magneto-centrifugal picture
\citep{Blandford1982}, the initially rigid magnetic field lines must be
inclined away from the disk normal (i.e., the jet axis) direction by more
than $30^\circ$ in order to {\it centrifugally} launch the outflow from the
disk surface and accelerate it to a high speed.  This requirement means that
the magneto-centrifugally driven outflow must start out as a wide-angle wind
with a range of flow directions
\citep{Shu1994,Ferreira1997,Krasnopolsky2003}.  As the wind propagates to
large distances,  inner streamlines are gradually collimated by the
``hoop stresses" associated with the toroidal magnetic field into a
high-density jet along the axis, while the remaining part remains
wide-angled (Shu et al.  1995, Shu1995 hereafter; Krasnopolsky et al. 
2003).  The slow collimation leads to two generic predictions for the
magneto-centrifugally driven jets: (1) the jet material near the base moves
in a wider range of radial directions (referred to as ``radial flow" hereafter) than
farther out, and (2) the jet is always surrounded by a (more tenuous)
wide-angle wind.  Here we report new observations of HH 212 to test these
predictions.

\section{Observations} 

Observations of the HH 212 protostellar system were obtained with Atacama
Large Millimeter/submillimeter Array (ALMA) in Band 7 centered at a
frequency of $\sim$ 341.5 GHz on 2017 November 27 in Cycle 5
(2017.1.00044.S).  The target was observed with a total time of $\sim$ 98
minutes and projected baselines of $\sim$ 60$-$8500 m.  Four spectral
windows were used, with three in continuum mode having a bandwidth of 2 GHz
and one in spectral mode having a bandwidth of 1.875 GHz.  The SiO line was
included in the spectral mode with a velocity resolution of $\sim$ 1.69
\vkm{} per channel.  CASA version 5.1 was used to calibrate the visibility
data and generate the SiO line data with continuum subtracted.  We used a
robust factor of 0.5 for the visibility weighting to generate SiO channel
maps with a synthesized beam of \arcsa{0}{036}$\times$\arcsa{0}{032} (or
14.4 au$\times$12.8 au) at a position angle of $\sim$ $-$78\degree{} and a
noise level of $\sim$ 0.75 \mJyb{} (or $\sim$ 7 K) in each channel.  These
channel maps are then used to generate the total intensity map and
position-velocity diagrams.

\section{Results}

\subsection{SiO Observations of the HH 212 Jet}

Figure \ref{fig:SiOPV} shows the inner part of this jet within 1400 au of
the protostar in SiO J=8-7 line.  The jet is well detected in SiO within
$\sim$ 200 au of the protostar, surrounded by the SO shells (blue dotted
curves) detected before \citep{Lee2021DW}.  It is highly collimated and
appears cylindrical (though knotty), extending out from the disk.  No SiO
jet emission can be seen down to the protostar because it is blocked by the
optically thick disk \citep{Lee2017Disk}.  In the south further away, some
faint clumpy emission is also seen along the jet axis, tracing the jet
itself.  In the north further away, a chain of roughly equally spaced wide
and flat structures are seen along the jet axis, each one oriented almost
perpendicular to the jet axis, probably tracing the internal shocks. 
Position-velocity (PV) diagram of the SiO emission along the jet axis shows
that the radial velocity range (i.e., line width) of the collimated jet
along the line of sight (LOS) increases toward the protostar, with its
Gaussian FWHM line width increasing from 5$-$7 \vkm{} at 160 au to 16$-$20
\vkm{} at 20 au.  It indicates that, despite the cylindrical appearance of
the observed jet, the outflowing material in the jet at the base is moving
more radially outward from the innermost part of the disk, consistent with
magneto-centrifugal models (see e.g., Figure \ref{fig:SiOPV}d).


A similar increase of line width towards the source has been detected
towards more evolved T Tauri jets, e.g., DG Tau and RW Aur
\citep{Pyo2003,Pyo2006}.  In those cases, the line profiles become more
complex towards the sources, showing prominent low-velocity emission
referred to as the low-velocity component (LVC).  This LVC also shows a
decrease in radial velocity towards the source, e.g., in DG Tau
\citep{Pyo2003,Maurri2014}.  The observed line width of the RW Aur jet near
the source was first found to be too large to be fully explained by a
divergent constant velocity flow of 4\degree{} in opening angle
\citep{Hartigan2009}, but later reproduced with an X-wind \citep{Liu2012}. 
It is thus not yet clear whether part or all of the LVC traces the
wide-angle base of the collimated jet or whether it traces different flow
components, e.g., an outer disk wind \citep{Simon2016,Nisini2018} or
entrained matter.


In the case of HH 212, no clear LVC can be identified in the PV structure. 
Although the peak emission of knots N1 and N2 shows a
decrease of radial velocity towards the source, the line wing emission of
these knots extends to higher and lower radial velocities, producing broader
line profiles towards the source. The jet interaction
with the SO shell at the base could contribute to the widening of the line
profiles close to the source on the low-velocity side, since the SO shell
has a low velocity near the base (see Section \ref{sec:shell}).


\subsection{Velocity and Long-range Acceleration of the HH 212 Jet}


Measurements of jet velocity can provide an additional constraint to the
models.  Figure \ref{fig:properm} shows the jet velocity at a distance from
$\sim$ 40 to 160,000 au, derived from the proper motion of the SiO jet
measured here from $\sim$ 100 to 1000 au (see Appendix) and those previously
reported at other distances along the jet axis
\citep{Claussen1998,Lee2015,Reipurth2019}.  Panel a shows the velocities at
the measured positions while panel b shows the velocities at the expected
positions at the date of current observations assuming ballistic motion.  In
either case, an apparent acceleration is seen over a large distance, with
the velocity increasing from $\sim$ 50$-$60 \vkm{} at 40$-$50 au, to $\sim$
90 \vkm{} at 1000 au, and then to $\sim$ 140 \vkm{} at 3,000 au.  The
velocity further away remains roughly constant.  Current MHD ejection model
might have difficulty in producing such a long-range acceleration.  Most MHD
ejection models predict that terminal jet velocities are reached within a
distance of $z = 100 r_0$, where $r_0$ is the anchoring radius of the
streamline in the disk \citep{Jacquemin-Ide2019,Tabone2020}.  Therefore,
further work is needed to check if other effects such as time variability in
the ejection velocity and geometrical projection can produce this apparent
acceleration.

\subsection{Modeling the HH 212 Jet}



As discussed earlier, radial flow is a generic feature of the
magneto-centrifugal wind at large distances from the (small) launching
region.  This asymptotic behavior is not only valid for an X-wind
(Shu1995), but also for an inner disk wind \citep{Cabrit1999,Pudritz2007},
especially if mass loading is concentrated towards the inner disk edge
\citep{Krasnopolsky2003}.  In the X-wind, the gas speed along all streamlines
is roughly similar while in the disk wind, the flow speed decreases as the
anchoring radius of the streamline increases.  The dense core of these winds
appears as a collimated jet.  For a jet almost in the plane of the sky, the
line profile is expected to widen towards the source to both lower and
higher velocities, in both the X-wind and disk wind.




We will focus our comparison to the asymptotic X-wind (Shu1995), for
which analytic solutions exist.  The wind solution is governed by the
magnetic field to mass flux ratio $\beta$ and the amount of specific angular
momentum $J$ carried by the gas and field along the streamlines.  It is an
asymptotically radial wind, but with the central part collimated by the
tension force of the toroidal magnetic field into a dense jet.  To be
specific, we adopt the X-wind parameters reported in Shu1995, with the mean
values averaged over the streamlines $\bar{J} \sim 3.7053$ and $\bar{\beta}
\sim 1.1675$.  This $\bar{J}$ value is similar to the mean value previously
found in young protostellar jets \citep{Lee2020Jet}.  It determines the
terminal velocity and specific angular momentum of the jet to be $\sim
v_{k,x} \sqrt{2 \bar{J}-3}$ and $\sim \bar{J}\; l_{k,x}$, respectively,
where $v_{k,x}$ and $l_{k,x}$ are the Keplerian rotation velocity and
specific angular momentum at the launching point called ``the X-point".

%





This asymptotic X-wind model is self-similar and scaled with the mass of the
protostar $M_\ast$, the radius of the launching point $R_x$, and the
mass-loss rate of the wind $\dot{M}_w$.  Since the total mass of the
protostar and the disk was estimated to be $\sim 0.25\pm0.05$ \solarmass{}
\citep{Lee2017COM} and the disk itself was estimated to have a mean mass of
0.10$\pm0.05$ \solarmass{} \citep{Lee2017Disk,Lee2021Pdisk}, we have $M_\ast
\sim 0.15\pm0.07$ \solarmass{}.  Then with the adopted $\bar{J}$ and the
observed specific angular momentum of $\sim$ 10$\pm3$ au \vkm{}
\citep{Lee2017Jet}, we have $l_{k,x}\sim$ 2.7$\pm0.8$ au \vkm{}, and thus
$R_x \sim 0.05\pm0.02$ au and $v_{k,x}\sim 51.7\pm15.9$ \vkm{}.  This model
produces a terminal jet velocity of $\sim$ 109$\pm33$ \vkm{}, similar to the
mean velocity of the jet and consistent with the jet velocity measured at
large distances within the errorbars (see Figure \ref{fig:properm}).  The
mass-loss rate in the jet was previously estimated to be
$\sim(1.0\pm0.5)\times 10^{-6}$ \msperyr{} at a large distance of $\sim
4000-8000$ au \citep{Lee2015}.  Considering that the mass-loss rate in the
wide-angle component of the X-wind can still be comparable to that in the
jet at that large distance (Shu1995), we assume $\dot{M}_w \sim
(1.5\pm0.7)\times10^{-6}$ \msperyr{}.  The resulting wind velocity, rotation
velocity, and density have been given in Shu1995.  The jet is the densest
core of the wind, with $\beta \sim 1$ and $J \sim 3.2$ within 1000 au of the
protostar (see Figure 1 in Shu1995).  Using Equation 4b in Shu1995 with
these $\beta$ and $J$ values, we find that the jet velocity in the model
increases slowly from 91$\pm28$ to $93\pm29$ \vkm{} at a distance from 40
and 1000 au.  Although the jet velocity at 1000 au in the model is
consistent with our measurement, the jet velocity at 40 to 300 au is about
two times the observed values there (see Figure \ref{fig:properm}).  The
actual X-wind could have a slightly different acceleration efficiency and
detailed numerical calculations \citep{Najita1994} are needed to fit the
observations more closely.

\begin{figure}
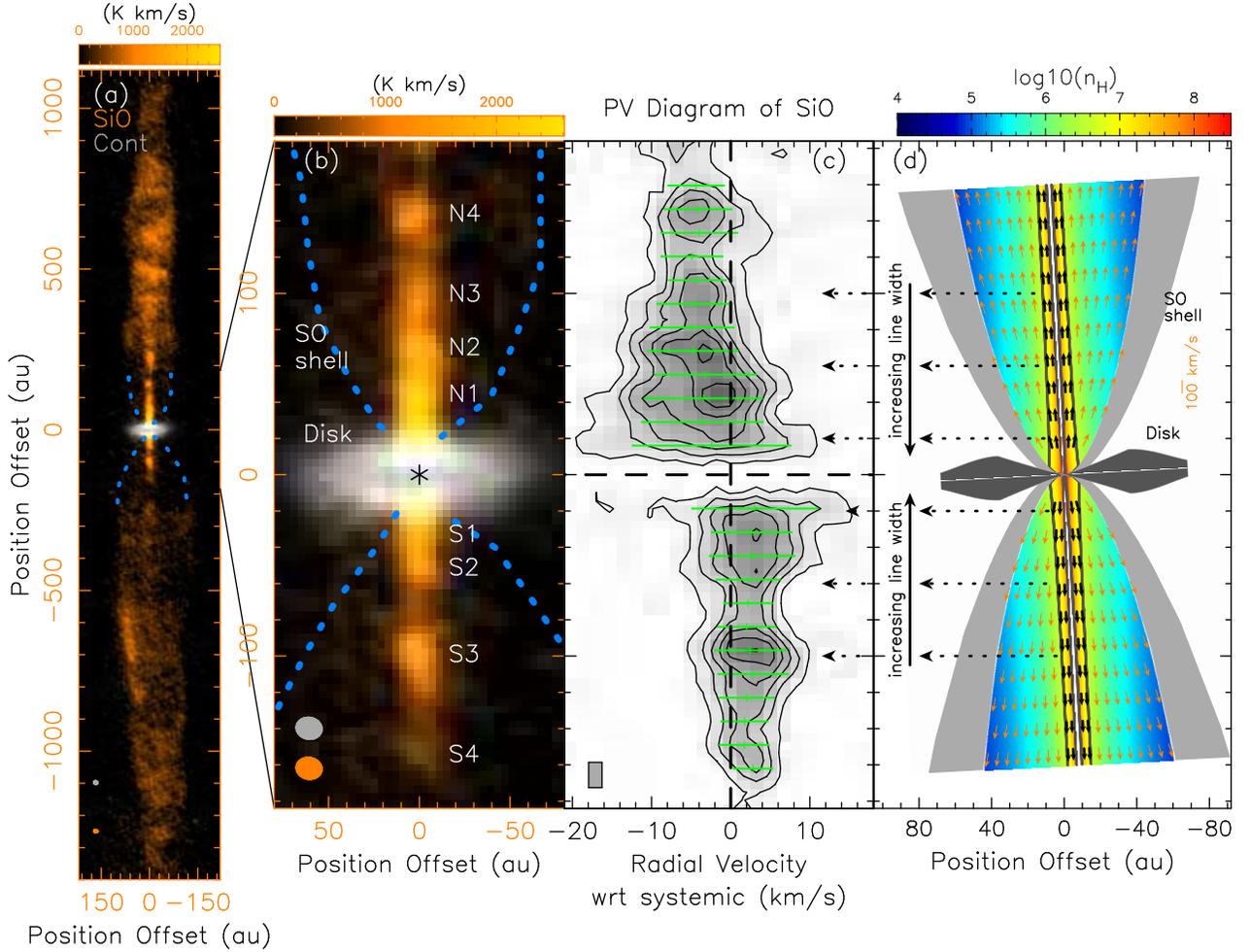
 \centering 
\putfig{0.7}{270}{f1.eps} 
\figcaption[]
{The HH 212 jet and model:  \tlabel{a} shows the SiO total intensity
map (orange) of the jet within 1400 au of the protostar at
\arcsa{0}{036}$\times$\arcsa{0}{032} resolution integrated from $-$16.1
to 17.6 \vkm{}, with the 350 GHz continuum map (gray, adopted from
\citet{Lee2021DW}) of the dusty disk.  They are rotated by 22.5\degree{}
clockwise to align the jet axis vertically.  The dotted blue curves show the
shells previously detected in SO \citep{Lee2021DW}.  \tlabel{b} shows the
innermost jet within $\sim$ 200 au of the protostar, with the nomenclature
of the knots from \citet{Lee2017Jet}.  \tlabel{c} shows the PV diagram along
the jet axis for the innermost jet,  made with a transverse width of 8
au.  The contours start at 3$\sigma$ with a step of 4.5$\sigma$, where
$\sigma=5.8$ K.  The green lines show the FWHM line widths
obtained from Gaussian fit. \tlabel{d} shows the number density and
velocity in our X-wind model, with the dusty disk (dark gray) and shells
(light gray), illustrating the increase of SiO line width of the jet towards
the protostar.  Here the jet is the densest core of the wind outlined by the
gray curves.
\label{fig:SiOPV}}
\end{figure}

\begin{figure}
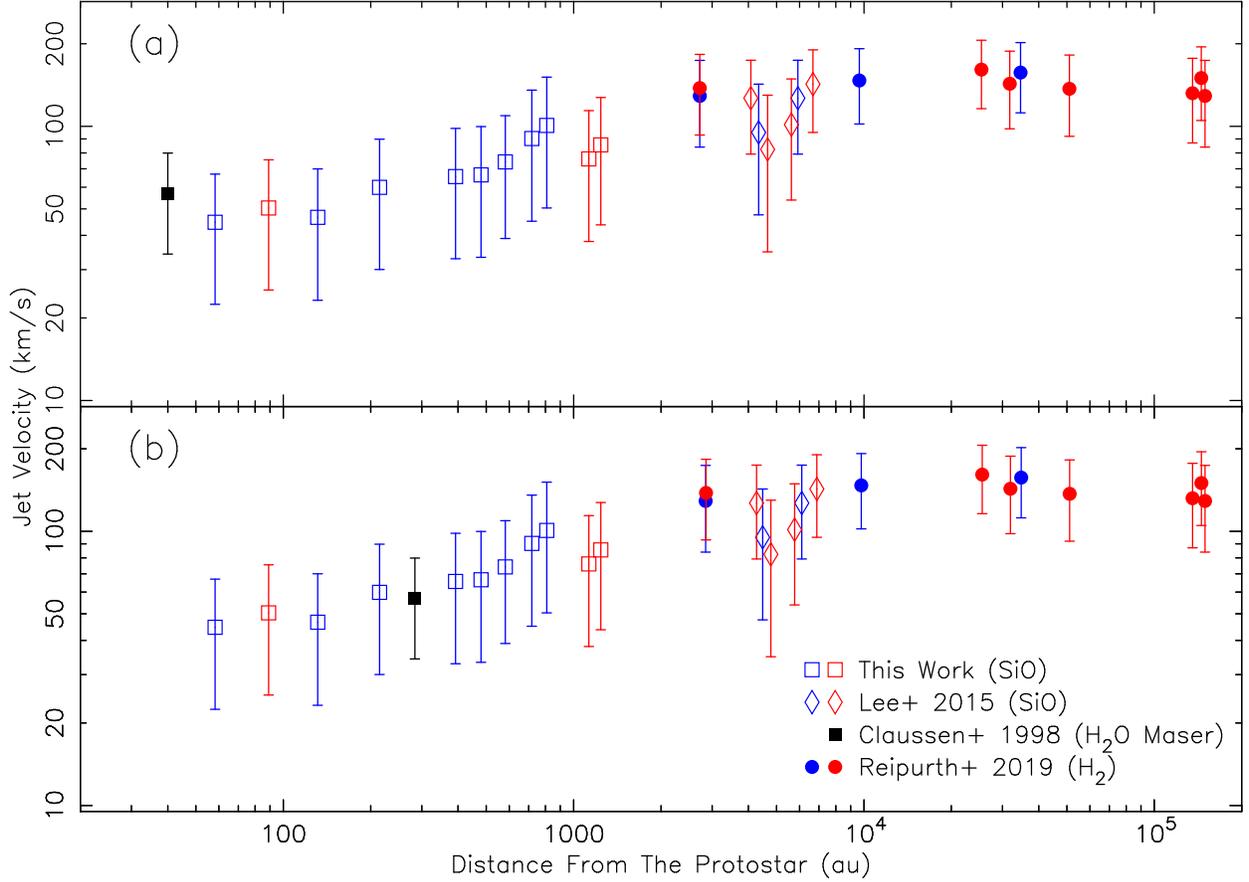
 \centering 
\putfig{0.68}{270}{f2.eps} 
\figcaption[]
{Jet velocity of HH 212 measured at a distance from $\sim$ 40 to 160,000 au.
They are derived from the proper motion
of the SiO jet measured here from $\sim$ 100 to 1000 au (see Appendix) and
those previously reported at other distances along the jet
axis \citep{Claussen1998,Lee2015,Reipurth2019}.
Red and blue data points are for the redshifted and blueshifted jet components, respectively.
Since the jet is almost in the plane of the sky, 
the jet velocity is the same as the proper motion.
In (b), the data points are shifted to their expected positions at
the mean date (2016 Nov) of the new ALMA observations, assuming ballistic motion.
\label{fig:properm}}
\end{figure}

We can compare the model quantitatively to the observed SiO jet up to a
distance of $\sim$ 160 au, where no clear internal shocks are seen affecting
the structure and kinematics.  Since we do not intend to reproduce the
observed knotty structure in the jet, which could be due to episodic
ejections, the X-wind is assumed to be steady.  For this comparison, we
adopt $M_\ast \sim 0.15$ \solarmass{}, $R_x \sim 0.05$ au, and $\dot{M}_w
\sim 1.5\times10^{-6}$ \msperyr{}, with the resulting number density and
velocity shown in Figure \ref{fig:SiOPV}d.  To mimic the observations,
we also include a dusty disk (dark gray) to block the innermost jet and
shells (light gray) to set the outer boundary of the X-wind in the model.
 SiO molecules are assumed to have a temperature of 400 K, as in another
young SiO jet HH 211 \citep{Hirano2006}.  A radiative transfer code
\citep{Lee2021DW} adding the SiO line with an assumption of LTE is used to
calculate the model SiO maps and then the model PV diagrams.  The
inclination angles of the jet are required to be 3\degree{} and 1.8\degree{}
to the plane of the sky for the northern and southern components,
respectively, to match the observed mean jet velocity there.  This
difference in the inclination angles is acceptable, because the jet seems to
have a bending of 2\degree{}$\pm$1\degree{} \citep{Lee2007}.  Also, as
discussed earlier, since the observed jet velocity appears to be a half of
that in the current model, the inclination angles could be two times as
high, and become roughly the same as those found for the shells later. 
Without understanding the long-range acceleration, we postpone the modeling
of the jet with such inclination angles to a future investigation.

\begin{figure}
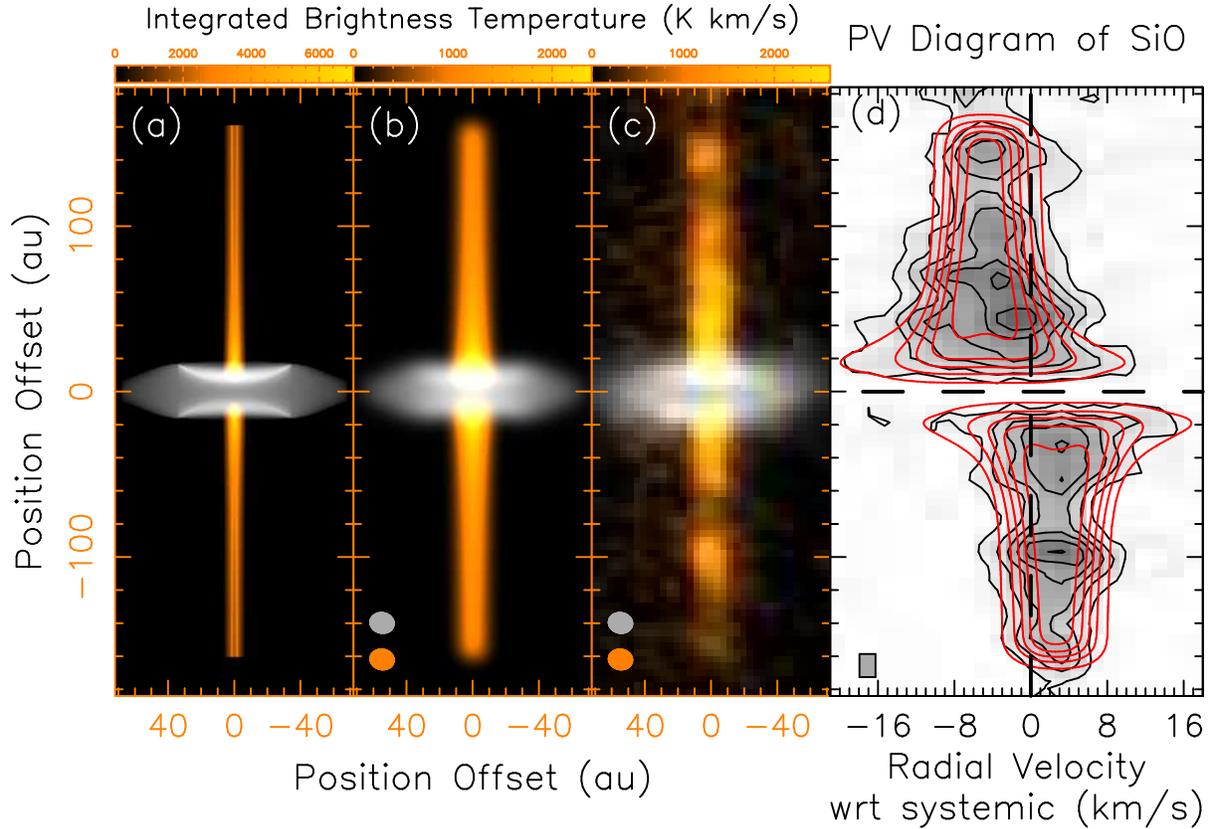
 \centering 
\putfig{1}{0}{f3.eps} 
\figcaption[]
{Comparison of the X-wind model and the observations of the HH 212
jet. \tlabel{a} shows the SiO map of the jet and the 350 GHz continuum
map of the disk derived from the model.  \tlabel{b} shows the map of the
jet and disk convolved to the observed resolution.  \tlabel{c} shows the
observed SiO jet and dusty disk.  \tlabel{d} shows the comparison of the PV
structures between the model (red contours) and observations (gray image
with black contours). 
Contour levels are the same as those in Figure \ref{fig:SiOPV}c.
\label{fig:modeljetpv}}
\end{figure}
\begin{figure}
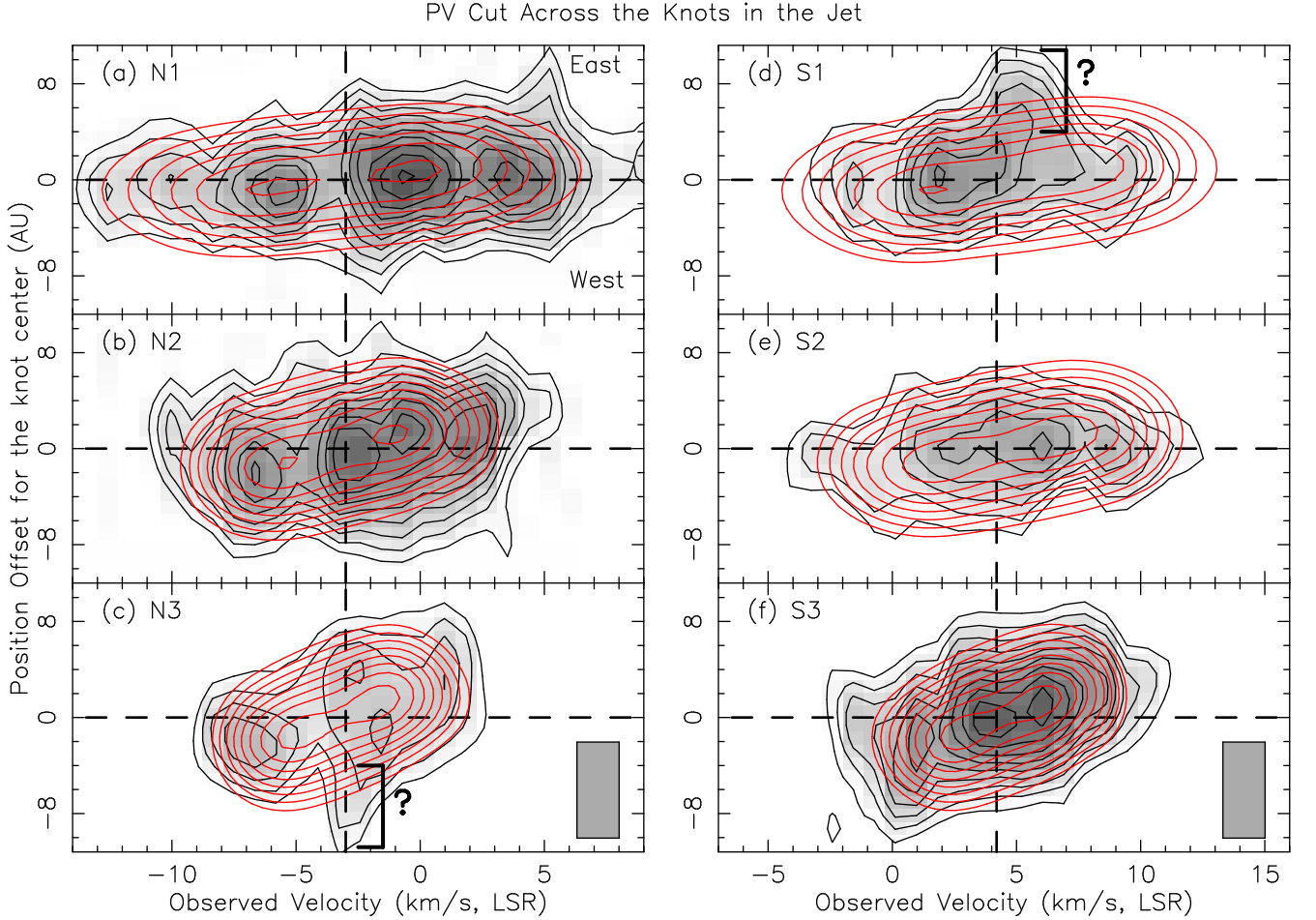
 \centering 
\putfig{0.7}{270}{f4.eps} 
\figcaption[]
{Comparison of the velocity gradients in the PV structures across the
knots in the innermost jet due to jet rotation. The observed PV structures 
(black contours with gray image) were adopted from
\citet{Lee2017Jet} at 8 au spatial resolution and 1.7 \vkm{} velocity resolution.
The model PV structures (red contours) are derived from the X-wind model at the same resolutions. 
The contour levels start from 4$\sigma$ with a step of 1$\sigma$, where $\sigma$ = 21.3 K. 
\label{fig:modelrotation}}
\end{figure}

In order to match the observed SiO emission intensity in the jet, the SiO
abundance is required to be [SiO/H]$\sim 3.3\times10^{-6}$, roughly
consistent with the molecular synthesis model \citep{Glassgold1991} for a
collimated fast wind similar to the X-wind.  In addition, in order to match
the observed width and velocity range at the base of the jet, the SiO
abundance is set to zero when the number density (in H) drops below
$6\times10^6$ \cmc{}, as marked by the gray curves in Figure
\ref{fig:SiOPV}d.  This density could be considered either as the minimum
value required to excite the SiO J=8-7 emission (which has a critical
density of $\sim 2.5\times10^7$ \cmc{} \citep{Lee2021DW}), or as the minimum
value where the SiO molecules have formed.  As seen in Figure
\ref{fig:modeljetpv}, the model produces a hollow SiO jet and can roughly
reproduce the observed width of the SiO jet after convolution to the
observed resolution.  This model can also reproduce the PV structure along
the jet axis (see Figure \ref{fig:modeljetpv}d), with the Gaussian FWHM
line width increasing from $\sim$ 6 \vkm{} at 160 au towards the protostar
to $\sim$ 18 \vkm{} at 20 au, similar to the observations. Moreover, as
shown in Figure \ref{fig:modelrotation}, this model can also roughly
reproduce the transverse velocity gradients of the knots in the jet due to
jet rotation previously obtained at higher resolution \citep{Lee2017Jet}. 
This match further supports the jet to be launched by magneto-centrifugal
force and the jet indeed carries away roughly the same amount of specific
angular momentum as in the model.



\subsection{Modeling the Shells Around the HH 212 Jet}\label{sec:shell}



\begin{figure}
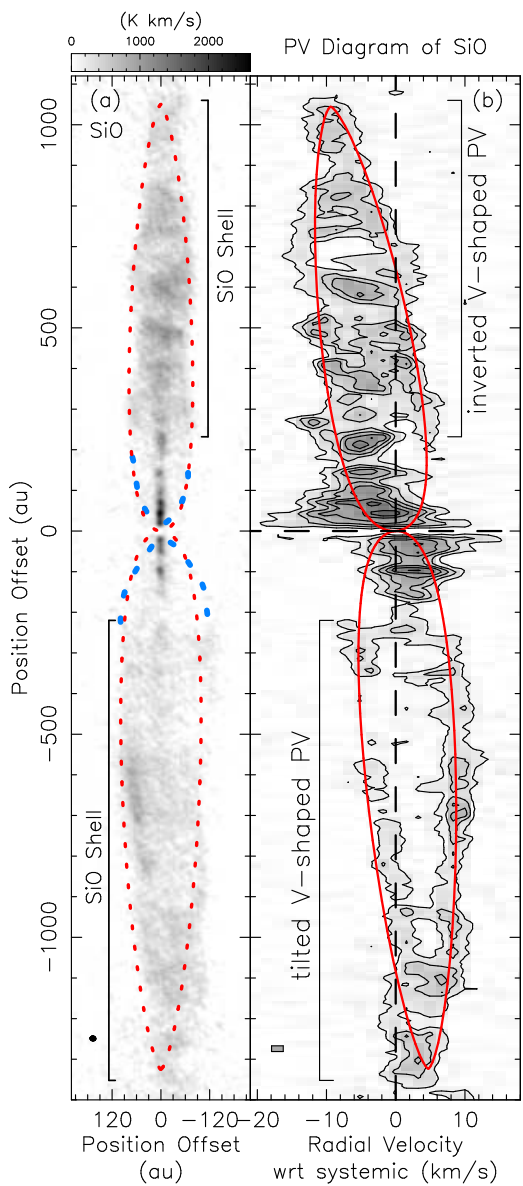
 \centering 
\putfig{0.6}{0}{f5.eps} 
\figcaption[]
{Comparison of the shell structure and PV structure along the jet axis between the model and observations.
The model structure (red dotted curves) and the model
PV structure (red solid curves) of the shells are derived from an interaction of the X-wind with an extended
disk wind. The dotted blue curves outline the shells previously detected in SO \citep{Lee2021DW}.
The contours start at 3$\sigma$ with a step of 4.5$\sigma$, where $\sigma=5.8$ K.
\label{fig:SiOshell}}
\end{figure}

Faint hollow shells are also seen in SiO in the north and south surrounding the jet
axis extending out from 200-300 au to more than 1000 au, as shown in
Figure \ref{fig:SiOshell}.  They connect to those previously seen in SO at
the base back to the disk (blue dotted curves), forming lobe structures
around the jet axis.  The PV diagram of the SiO emission along the jet axis
shows a tilted V-shaped PV structure arising from the shell in the south,
and an inverted V-shaped PV structure arising from the shell in the north. 
These V-shaped PV structures and the previously detected parabolic PV
structures of the SO shells near the base \citep{Lee2018DW} form tilted
lobe-shaped PV structures.



Previously, the shells were thought to be produced by a pure collimated
jet through jet-driven bow shocks at the tips of the shells \citep{Lee2021DW}. 
However, the spur PV structures, which are the signatures of the jet-driven
bow shocks and have a large velocity range at the bow tips due to sideways
ejection \citep{Lee2001,Tabone2018}, are not seen here in the PV diagram at
the tips of the shells.  Instead, the shells could be driven by X-winds that
have wide-angle radial components, because tilted lobe-shaped PV structures
have been modeled before with such winds to account for the large-scale
molecular outflows around the jets \citep{Shu1991,Lee2001,Shang2020}.

Here we construct a momentum-driven shell model to account for the structure
and kinematics of the shells.  In this model, the shells are produced as the
X-wind expands into the extended disk wind coming out from the outer disk
detected before in SO \citep{Tabone2020,Lee2021DW}.  This extended disk wind
is a hollow wind launched from the outer disk with a radius from $\sim$ 4 to
40 au.  It is magnetic and vanishes inside 4 au, probably 
because of a decrease of disk magnetization \citep{Lee2021DW}. It has a
mass-loss rate of $\sim 10^{-6}$ \msperyr{}, with the number density
decreasing with the increasing height from the disk
midplane approximately as (in spherical coordinates) \begin{equation} n_d
\sim 6\times10^6
 \Big(\frac{\sin \theta}{\sin 45^\circ}\Big)
 \Big(\frac{100\;\textrm{au}}{r\cos\theta}\Big)^p \;\cmce 
\end{equation}
where the power-law index $p=4/3$.  Here the sine factor is to make the
wind hollow, with $\theta$ being the angle measured from the jet axis. 
Since the SO emission derived from this wind decreases with height slower
than that detected in the observations \citep{Lee2021DW}, the actual density
of the wind could decrease faster with height.  Therefore, we increase
from $p=4/3$ to $p=2$, which also simplifies our
calculation of the shells later. 
The velocity of the extended disk wind is much smaller than
that of the X-wind and thus ignored in our calculation.
According to \citet{Lee2001}, the resulting shell velocity is
\begin{equation} 
v_s = \frac{v_w}{1+\eta^{-1/2}} 
\end{equation} 
where $v_w$ is the velocity of the X-wind
and $\eta = n_w/n_d$, with $n_w$ and $n_d$ being the number density of the X-wind and disk wind, respectively.
At the tip of the shell, $\eta \gg 1$ and thus $v_s \sim v_w$.
The resulting shell has a lobe-like structure
with
\begin{equation}
r_s \approx \frac{L}{1+\eta^{-1/2}}
\end{equation}
where $L$ is length of the lobe.





Figure \ref{fig:SiOshell} shows the model structure in red dotted curves and
model PV structure in red solid curves, with $v_w \sim 90$ \vkm{} and $L
\sim 1050$ au in the north and 1350 au in the south, as measured from the
observations.  The shells have a dynamical age of $\sim L/v_w$, the
same as their associated jet component, about $\sim$ 60 yrs. The
inclination angle of the shell to the plane of sky is $\sim$ 6\degree{} in
the north and 3\degree{} in the south.   Like the jet, the small
difference of inclination angles between the two sides is acceptable. As
can be seen, this model reproduces the shell structure and kinematics
reasonably well.  Moreover, in this model, the shell velocity increases to
$\sim$ 20-25 \vkm{} at a distance of $\sim$ 300 au from the protostar, high
enough to sputter Si or SiO from the grains in the dusty disk wind and form
the SiO in the gas phase \citep{Schilke1997}, producing the observed SiO
emission in the shells.







\section{Summary and Discussion}

Regardless of the uncertainties in the model parameters and measurements,
the simple X-wind model can still reproduce the key features (including
radial flow, rotation, and structure) of the jet as the densest core
of the wind and the key features (including structure and kinematics) of the
shells with the interaction between the wide-angle radial component of the
wind and an extended dusty disk wind.  It is expected that an inner
disk wind can also reproduce the key features.  Thus, the jet is likely a
dense core of either an X-wind or an inner disk wind.









Interestingly, the wide-angle radial component, if does exist,
can also solve other existing problems.  In particular, since a pure
collimated jet has difficulty in producing the observed width and kinematics
of large-scale molecular outflows \citep{Lee2002}, the wide-angle radial
component can provide a promising solution to the width and kinematic
problems of those outflows \citep{Lee2000,Shang2006,Arce2007,Zhang2019}. 
Moreover, it is still unclear why chondritic materials and Ca-Al-rich
inclusions are detected throughout the Solar System.  It could be that they
were originally produced during the star formation in the innermost disk
where the temperature was high enough to form them, but later flung out by
the wide-angle wind component to rain down everywhere in the disk
\citep{Shang2000}.

\acknowledgements
We thank the anonymous referee for insightful comments.
This paper makes use of the following ALMA data:
ADS/JAO.ALMA\#2017.1.00044.S.  ALMA is a partnership of ESO (representing
its member states), NSF (USA) and NINS (Japan), together with NRC (Canada),
NSC and ASIAA (Taiwan), and KASI (Republic of Korea), in cooperation with
the Republic of Chile.  The Joint ALMA Observatory is operated by ESO,
AUI/NRAO and NAOJ.  C.-F.L.  acknowledges grants from the Ministry of
Science and Technology of Taiwan (MoST 107-2119-M-001-040-MY3,
110-2112-M-001-021-MY3) and the Academia Sinica (Investigator Award
AS-IA-108-M01).  Z.-Y.L.  is supported in part by NASA 80NSSC20K0533 and NSF
AST-1910106.

\clearpage

\appendix
\setcounter{figure}{0}
\renewcommand{\thefigure}{A\arabic{figure}}

\section{Proper Motion}

\begin{figure}
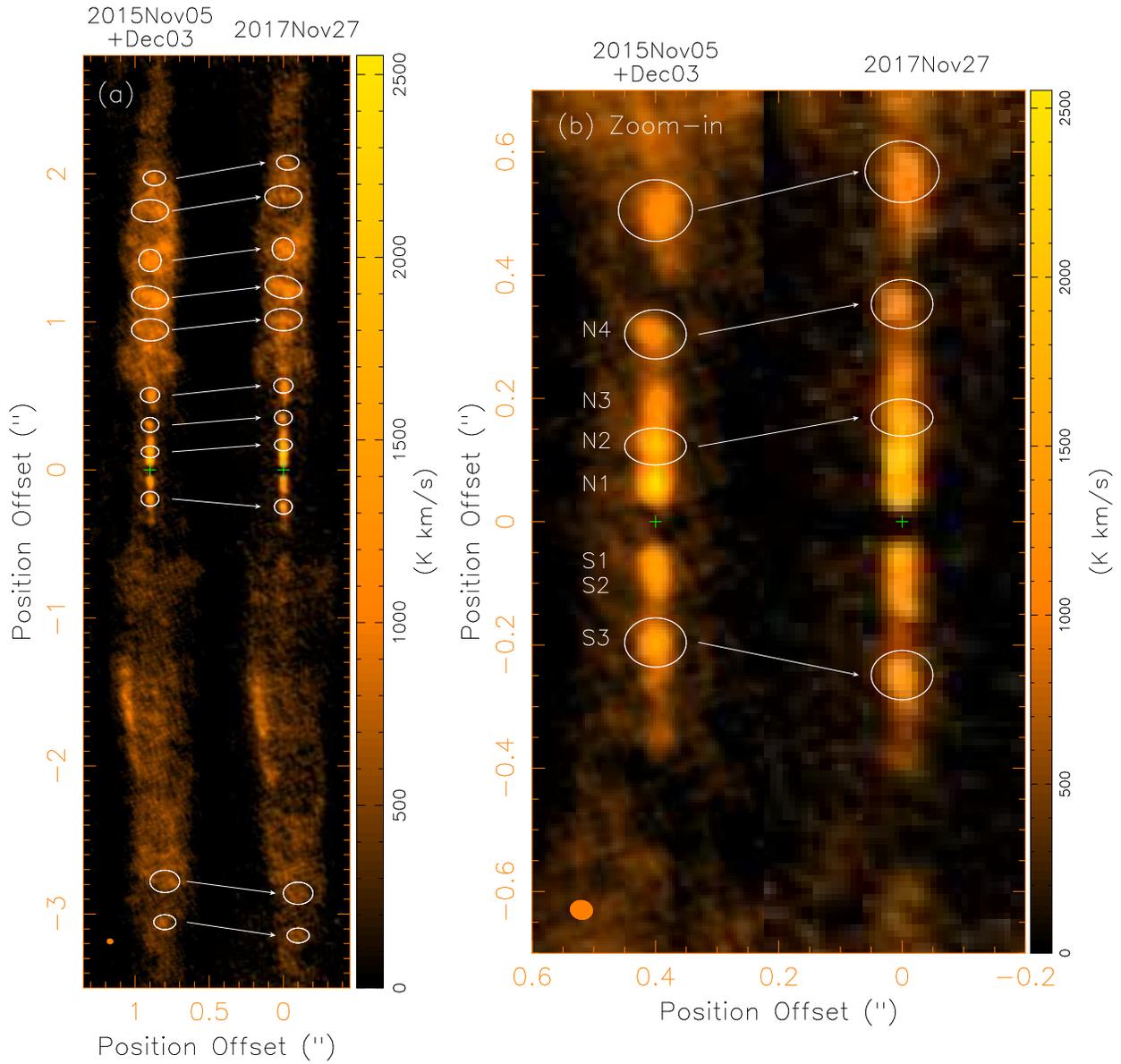
 \centering 
\putfig{1.0}{0}{fA1.eps} 
\figcaption[]
{Proper motion measurement of the SiO jet by comparing the SiO
total intensity maps in two epochs at
\arcsa{0}{036}$\times$\arcsa{0}{032} resolution.  The white arrows show the position
shifts of the isolated knots and internal shocks (marked with white ellipses)
between the two epochs.
\label{fig:SiOpm}}
\end{figure}

Proper motion of the jet can be measured by comparing the SiO total
intensity map of the jet presented in this paper to that obtained about 2
years earlier on 2015 November 05 and December 03 reported in
\citet{Lee2017Jet}.  The SiO map taken earlier has been convolved to have
the same angular resolution as that here.  As shown in Figure
\ref{fig:SiOpm}, the jet shows roughly the same structures in the two
epochs.  The white arrows show the position shifts of the isolated knots and
internal shocks of the jet between the two epochs.   The inner knots are
not spatially well resolved, and only knots N2, N4, and S3 can be roughly
separated to measure the proper motion (see Figure \ref{fig:SiOpm}b for the
zoomin). The position shifts are found to range from \arcsa{0}{05} to
\arcsa{0}{11}, which is $\sim$ 1.5 to 3 synthesized beams.  Due to the
cooling, expansion, and other factors, the emission features from the same
gas could be different between the two epochs.  Hence the uncertainties of
the measurements are assumed to be a half of the position shifts.  The
proper motion can then be obtained by dividing the position shifts by the
duration of 2 years.

\end{document}